\documentclass{elsarticle}

\usepackage{graphicx}
\usepackage{bm}
\usepackage{cancel}
\usepackage{color}
\usepackage[utf8]{inputenc}
\usepackage{amsmath}
\usepackage{amssymb}

\begin{document}
\bibliographystyle{plain}

\begin{frontmatter}

\title{Dark energy based on exotic statistics}

\author[1,2]{M. Hoyuelos}
\ead{hoyuelos@mdp.edu.ar}
\author[2,3]{P. Sisterna \corref{cor1}}
\ead{sisterna@mdp.edu.ar}
\cortext[cor1]{Corresponding author}

\affiliation[1]{Instituto de Investigaciones Físicas de Mar del Plata (IFIMAR -- CONICET), Funes 3350, 7600 Mar del Plata, Argentina}
\affiliation[2]{Departamento de Física, Facultad de Ciencias Exactas y Naturales, Universidad Nacional de Mar del Plata, Funes 3350, 7600 Mar del Plata, Argentina}
\affiliation[3]{Departamento de Filosofía, Facultad de Humanidades, Universidad Nacional de Mar del Plata, Funes 3350, 7600 Mar del Plata, Argentina}

\begin{abstract}
Dark energy is an elusive concept, which has been introduced two decades ago in order to make the acceleration of the universe a comprehensible phenomenon. However, the nature of this energy is far from being understood, both from a fundamental as well as an observational way. In this work we study cosmological consequences of the existence of particles (which we called ``ewkons'' in a previous work) which are quasi distinguishable, obey unorthodox statistics, and have an equation of state similar to many existent dark energy candidates (including negative relation between pressure and energy density). We find an effective scalar field description of this ewkon fluid, and obtain cosmological solutions for the dark energy dominated epoch. This can be considered as a one-parameter class of dark energy models.
\end{abstract}


\end{frontmatter}

\section{Introduction}

The thermodynamic properties of macroscopic systems can be derived in a number of ways, including the study of the transition rates between possible states. Let us consider a system of non-interacting quantum particles in contact with a reservoir at temperature $T$ and chemical potential $\mu$. An expression for the transition rate between levels with energy $\epsilon_1$ and $\epsilon_2$, with $n_1$ and $n_2$ particles respectively, in terms of the residual chemical potential was recently derived \cite{hoyuelos3}. It was also shown that, if the transition rate depends on the number of particles in the destination level, then Fermi-Dirac (FD $+$) and Bose-Einstein (BE $-$) statistics are deduced, with the well known average occupation numbers  $\bar{n}_\pm = 1/(e^{(\epsilon-\mu)/T} \pm 1)$ (natural units with $c=\hbar=k_B=1$ are used). On the other hand, if time reversal is applied, then we expect that the transition rate will depend on the number of particles in the origin level and, in this case, ewkons ($+$) and genkons ($-$) statistics are obtained, with occupation numbers $\bar{n}_\pm = e^{-(\epsilon-\mu)/T} \pm 1$ (these exotic statistics where introduced in Ref.\ \cite{hoyuelos-sisterna}). The symmetry and simplicity of the result are features that encourage a more profound analysis.

The analysis of ewkon statistics turns out to be particularly interesting from the cosmological point of view. An ideal gas of ewkons has negative pressure. Furthermore, the barotropic parameter can be close to $-1$,  features that make ewkons appropriate to describe dark energy (for a review of quintessence models, see, for example, \cite{Tsujikawa2013}). Ewkon statistics was originally derived in Ref.\ \cite{hoyuelos-sisterna} from the assumption of free diffusion in energy space and the adjustment of an ``interpolation parameter''. Simpler conditions are required in the derivation of Ref.\ \cite{hoyuelos3} mentioned before.  Non-relativistic ewkons of mass $m$ and a massless scalar field of ewkons were analyzed in \cite{hoyuelos1} and \cite{hoyuelos2} respectively. Here we study the thermodynamic properties of ewkons throughout the history of the Universe assuming that their present density accounts for the bulk of the present density of dark energy. The purpose is to check the consistency of the hypothesis that dark energy has the statistics of ewkons.

Several authors have investigated the quantum formulation of particles with statistics different from BE or FD. A small sample of references is \cite{hartle,wilczek,tersoff,greenberg0,greenberg,bach,haldane,isakov,greenbergs,melic,dai} (see \cite{khare} for a review). 

The statistical properties of a system of non-interacting particles in contact with a reservoir at temperature $T$ and chemical potential $\mu$ are given by the grand partition function $\mathcal{Z}=\prod_\mathbf{k} \mathcal{Z}_\mathbf{k}$, where the sub-index refers to the mode with wave vector $\mathbf{k}$ and, using the base of number eigenstates, 
\begin{equation}\label{e.Z}
\mathcal{Z}_\mathbf{k} = \sum_{n} \delta_n\, e^{-n(\epsilon_\mathbf{k}-\mu)/T}
\end{equation}
is the grand partition function for particles that have energy $\epsilon_\mathbf{k}$; $\delta_n$ is the statistical weight factor. For Bose-Einstein (BE) statistics we have $\delta_n=1\;\forall n$, while for Fermi-Dirac (FD) statistics we have $\delta_0=\delta_1=1$ and $\delta_n=0$ for $n\ge 2$. Maxwell-Boltzmann statistics is obtained from $\delta_n=1/n!$. In order to calculate the grand partition function $\mathcal{Z}_\mathbf{k}$, the vacuum energy term $\epsilon_\mathbf{k}/2$ is removed as usual, for it leads to inconsistencies at the cosmological level (see for example \cite[p.\ 19]{kapusta}).

Statistical weights different from those of bosons or fermions may represent identical particles, albeit with some degree of distinguishability, that are outside the scope of the spin-statistics theorem (the spin-statistics connection applies to indistinguishable particles, see, for example, \cite[Ch.\ 4]{srednicki}; see \cite{huggett} for a clear distinction between the concepts of ``identicality'' and ``indistinguishability''). Such situations are not so rare at the fundamental level since, for example, two electrons with opposite spin can be treated as approximately distinguishable \cite[p.\ 315]{huggett}. In principle, quantum mechanics can be developed without the symmetrization postulate (that, in turn, implies the indistinguishability postulate), allowing more general statistics \cite{greenberg-comp,messiah-greenberg}.

The density and pressure of an ideal gas of ewkons are obtained from the corresponding partition function in Sec.\ \ref{s.parfun}, where large and small temperature regimes are discussed. Statistical effects can be represented by an effective potential. A scalar field effective description with its corresponding potential are introduced in Sec.\ \ref{s.potential}. The functional form of this potential is obtained for the dark energy dominated era. The dynamics of the scalar field is analyzed in Sec.\ \ref{s.scalar}. Sec.\ \ref{s.concl} contains the conclusions, where a brief comparison with present models of dark energy is included.

\section{Partition Function, density and pressure of ewkons}
\label{s.parfun}

Unlike bosons, the lowest energy state for ewkons is not $|0\rangle$, but $|1\rangle$. Then, we have
\begin{align}\label{e.Zewk0}
\mathcal{Z}_\mathbf{k} &= \sum_{n=1}^{\infty} \delta_n e^{-n(\epsilon_\mathbf{k}-\mu)/T} \nonumber\\
&= e^{-(\epsilon_\mathbf{k}-\mu)/T} \sum_{n'=0}^{\infty} \delta_{n'+1} e^{-n'(\epsilon_\mathbf{k}-\mu)/T},
\end{align}
where the substitution $n=n'+1$ was performed in the second line. We define the statistical weight, $\delta_{n'+1}$, as the Gibbs factor for distinguishable particles, that is $\delta_{n'+1}=1/n'!$, or, equivalently, $\delta_n = 1/(n-1)!$ \cite{hoyuelos2}. Then,
\begin{equation}\label{e.Zewk}
\mathcal{Z}_\mathbf{k} = \exp\left[ -(\epsilon_\mathbf{k}-\mu)/T + e^{-(\epsilon_\mathbf{k}-\mu)/T} \right].
\end{equation}
Therefore, the mean occupation number is
\begin{equation}\label{e.newk}
\bar{n}_\mathbf{k} = T\frac{\partial \ln \mathcal{Z}_\mathbf{k}}{\partial \mu} = e^{-(\epsilon_\mathbf{k}-\mu)/T} + 1.
\end{equation}
The main ingredients of the ewkon field definition are a lowest energy state other than the vacuum and a statistical weight related to the Gibbs factor. These ingredients lead to the number statistics of Eq.\ \eqref{e.newk}. The main motivations are that this number statistics can be deduced from simple assumptions on the transition rates \cite{hoyuelos3}, as mentioned in the introduction, and that the resulting thermodynamic properties have connections with dark energy, as shown in the next paragraphs.

We can now analyze the thermodynamic properties of an ideal gas of massless ewkons, or at least negligible rest energy compared with the kinetic energy ($\epsilon_\mathbf{k}=\sqrt{m^2 + k^2}\simeq k$), and zero chemical potential. The total grand partition function can be written as
\begin{align}
\frac{1}{V} \ln \mathcal{Z} &= \frac{1}{(2\pi)^3}\int {\rm d}\mathbf{k} \,g \ln\mathcal{Z}_\mathbf{k} \nonumber \\
&= \frac{1}{2\pi^2} \int_{0}^{\epsilon_{\rm m}} {\rm d}\epsilon \, g \epsilon^2 (e^{-\epsilon/T} - \epsilon/T)
\end{align}
where $g$ is a constant equal to the degeneracy, which we consider to lie between 1 and 10. Ewkons are taken as relativistic particles, with $\epsilon_\mathbf{k}=k$, and $k = |\mathbf{k}|$; sub-index $\mathbf{k}$ was removed in $\epsilon_\mathbf{k}$ for simplicity. 
We introduce a maximum energy $\epsilon_{\rm m}$ in order to avoid divergences; its value is fixed later using the energy conservation equation. The results for the energy density and the pressure are:
\begin{align}
\rho &= \frac{g}{2\pi^2} \int_{0}^{\epsilon_{\rm m}} {\rm d}\epsilon \,  \epsilon^3 \, \bar{n}_\mathbf{k} \label{e.rho} \nonumber \\
&=\frac{gT^4}{8\pi^2}\left[ (u^4+24)e^u - 4u^3 - 12u^2 - 24u - 24 \right]e^{-u}, \\
p &= \frac{T}{V} \ln \mathcal{Z} = - \frac{gT^4}{8\pi^2}\left[ (u^4-8)e^u + 4u^2 + 8u + 8 \right]e^{-u}, \label{e.p}
\end{align}
with $u=\epsilon_{\rm m}/T$. The equation of state or barotropic parameter is
\begin{equation}
w=\frac{p}{\rho}=-\frac{(u^4-8)e^u + 4u^2 + 8u + 8}{(u^4+24)e^u - 4u^3 - 12u^2 - 24u - 24}. \label{e.wu}
\end{equation}
Although this last expression is a quotient of quasi polynomials in $u$, it does not resemble any known dark energy parametrization, such as the Sendra-Lazkoz parametrization \cite{Sendra2012}, the Feng-Shen-Li-Li \cite{Feng2012}, Barboza-Alcaniz \cite{Barboza2008}, Chevallier-Polarski-Linder \cite{Chevallier2001,LinderPRL2003} or Jassal-Bagla-Padmanabhan \cite{Jassal2005} parametrizations, or even models with a Chaplygin like fluid \cite{Shenavar2020,Bento2004}.

In a homogeneous and isotropic universe, the momentum of each ewkon particle will decay as $a^{-1}$, where $a$ is the scale factor of the universe, so we have as usual $T\propto a^{-1}$; the subscript $0$ will denote the present epoch and we set $a_0=1$, so we can write $T=T_0/a$.

We consider a universe in which dark energy, with density $\rho_{\rm de}$ and pressure $p_{\rm de}$, behaves as ewkons. We also assume that there is no interaction with matter or radiation, so the energy conservation equation is 
\begin{equation}
\dot\rho_{\rm de}= - 3 \frac{\dot a}{a} (\rho_{\rm de}+p_{\rm de}) = - 3 \frac{\dot a}{a} (w+1) \rho_{\rm de}. \label{energy}
\end{equation}
Interactions may have been present during the very early stages of the universe, so it is reasonable to expect a value of $T_0$ similar to the present CMB (Cosmic Microwave Background) temperature, equal to $2.72548$ K \cite{fixsen}, or $2.34863\;10^{-4}$ eV.

As usual, adiabaticity is assumed: although $\rho_{\rm de}$ and $p_{\rm de}$ are time dependent, they can be calculated using equilibrium statistical mechanics. Then, using \eqref{e.rho} and \eqref{e.p} in the energy conservation equation \eqref{energy}, and taking into account that $T$ and $\epsilon_{\rm m}$ depend on time, after some algebra the following differential equation is obtained: 
\begin{equation}\label{e.em0}
(e^{\epsilon_{\rm m}/T}+1)\frac{\dot{\epsilon}_{\rm m}}{\epsilon_{\rm m}} = \left( 1 + \frac{T}{\epsilon_{\rm m}} \right) \frac{\dot{T}}{T},
\end{equation}
or, in terms of $u$ and $a$,
\begin{equation}
    \frac{(1+e^u)}{(1-u\, e^u)}\dot{u} = -\frac{\dot{a}}{a}.
\end{equation}
The solution is
\begin{equation}\label{e.em}
\frac{\epsilon_{\rm m}}{T} = \frac{\epsilon_\infty}{T} + e^{-\epsilon_{\rm m}/T},
\end{equation}
where $\epsilon_\infty$ is the (constant) value of $\epsilon_{\rm m}$ in the limit of small temperature, $T\ll \epsilon_\infty$, that corresponds to the limit when the scale factor (or time) diverges. See the Appendix for a more detailed derivation of \eqref{e.em}.

The limits of small and large temperatures are analyzed in the next subsections.

\subsection{Small temperature, $T\ll \epsilon_\infty$.}
\label{s.small}

In the limit $T\ll \epsilon_\infty$ we have $u\gg 1$ and $\epsilon_{\rm m} = \epsilon_\infty$. The density, pressure and barotropic parameter take the values
\begin{align}
\rho_{\rm de} &= \frac{g \epsilon_\infty^4}{8\pi^2} \label{e.rhosmallT} \\
p_{\rm de} &= -\frac{g\epsilon_\infty^4}{8\pi^2} \\
w &= -1.
\end{align}
It can be seen that, in this limit, ewkons behave as a cosmological constant that provokes the accelerated expansion of the Universe. Using the value for Hubble parameter obtained by the Planck Collaboration \cite{Planck2018}, the present total density (equal to the critical density assuming the Universe to be approximately spatially flat) is obtained from the Friedmann equation. Multiplying by the dark energy density parameter, $\Omega_{\rm de}\simeq 0.68$ \cite{Planck2018}, the present density of dark energy is $\rho_{\rm de}^0=2.53\;10^{-11}$ eV$^4$.   Assuming that we are currently in the small-temperature regime, we obtain that $\epsilon_\infty$ is equal to 0.0067 eV for $g=1$, or 0.0038 for $g=10$, that is between 16 to 28 times larger than $T_0$ (assuming $T_0$ to be the same as the CMB temperature). This difference is large enough to neglect the exponential term in Eq.\ \eqref{e.em}. In this regime $\rho_{\rm de}$ is constant and approximately equal to $\rho_{\rm de}^0$, which in turn is approximately equal to $\rho_{\rm de}^\infty$, so that
\begin{equation}\label{e.rho-inf}
    \rho_{\rm de}^\infty = \frac{g \epsilon_\infty^4}{8\pi^2}.
\end{equation}
Below we use $\rho_\infty$ instead of $\rho_{\rm de}^\infty$ to simplify the notation.

\subsection{Large temperature, $T\gg \epsilon_\infty$.}
\label{s.large}

For large temperatures, the first term in the right-hand side of Eq.\ \eqref{e.em} can be neglected, so we have
\begin{equation}\label{e.emlargeT}
\frac{\epsilon_{\rm m}}{T} =  e^{-\epsilon_{\rm m}/T},
\end{equation}
whose solution is $\epsilon_{\rm m} = 0.567\, T$ (note that the range of the variable $u$ lies between $u\rightarrow\infty$ for $t\rightarrow\infty$ to a constant value $u=0.567$ for early times). The corresponding values of density, pressure and barotropic parameter are:
\begin{align}
\rho_{\rm de} &= 2.15\;10^{-3}\, g T^4 \label{e.rholargeT}\\
p_{\rm de} &= 0.715\;10^{-3}\, g T^4 \label{e.plargeT} \\
w &= 1/3,
\end{align}
where the numbers in Eqs.\ \eqref{e.rholargeT} and \eqref{e.plargeT} can be computed with arbitrary precision. We can see then that in the large-temperature regime, ewkons behave as radiation, with $\rho_{\rm de} \sim a^{-4}$. 

From Eqs.\ \eqref{e.rhosmallT} and \eqref{e.rholargeT}, the scale factor at the crossover between both regimes is
\begin{equation}\label{e.atrans}
a_c = T_0 \left( \frac{2.15\;10^{-3} g}{\rho_0}\right)^{1/4}, 
\end{equation}
that corresponds to values between $0.02$ and $0.04$ ($49>z>24$) for $g$ between 1 and 10; the corresponding crossover temperature is between $0.01$ eV and $0.006$ eV ($115$ K and $70$ K).

\begin{figure}
	\includegraphics[width=\linewidth]{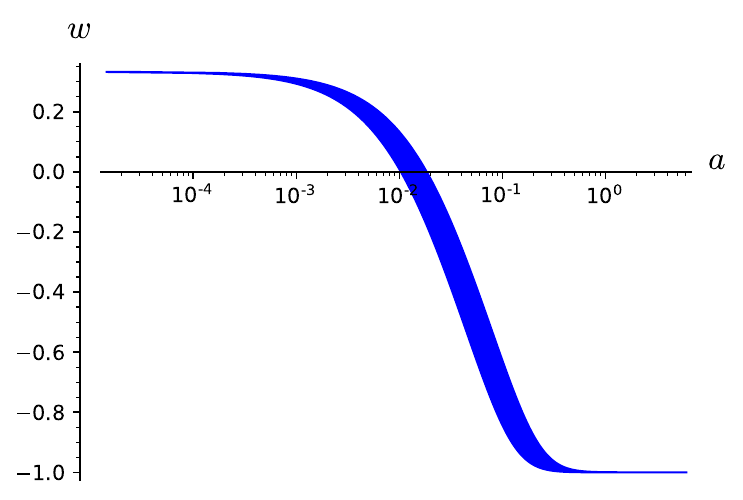}
	\caption{Barotropic parameter, $w$, for dark energy with ewkon statistics against the scale factor, $a$, in log scale. The curve corresponds to Eq.\ \eqref{e.wu}, where the value of $u=\epsilon_{\rm m}/T$ is obtained from Eq.\ \eqref{e.em} and $T=T_0/a$. The thickness variation of the curve represents values of the degeneracy, $g$, between 1 and 10. The parameter $w$ takes the asymptotic values $1/3$ and $-1$ for small $a$ and large $a$ respectively.} \label{f.w}
\end{figure}

\begin{figure}
	\includegraphics[width=\linewidth]{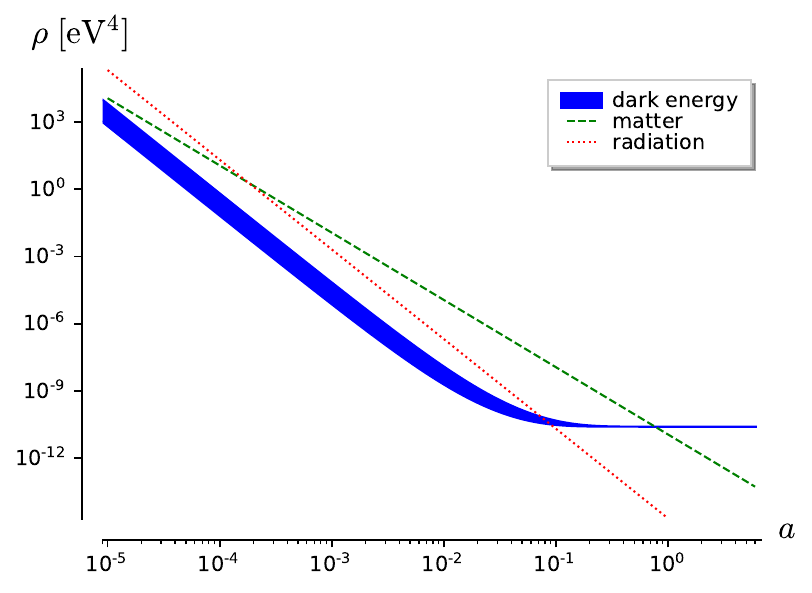}
	\caption{Density against scale factor in log-log scale. The density of dark energy (blue) is obtained from Eq.\ \eqref{e.rho} with $u$ calculated from Eq.\ \eqref{e.em}; the thickness of the curve correspond to values of $g$ between 1 and 10. Densities of matter (dashed line) and radiation (dotted line) are also shown for comparison; the parameters for these lines were taken from \cite[p.\ 30]{dimopoulos}.} \label{f.rho}
\end{figure}

The values of $w$ and $\rho_{\rm de}$ in both regimes against the scale factor are shown in Figures \ref{f.w} and \ref{f.rho} respectively. Also the densities of matter and radiation against $a$ are shown in Fig.\ \ref{f.rho} for comparison. Close to $a=1$, the density of ewkons, or dark energy, overcomes the density of matter and dominates at present. In the large temperature regime, when ewkons behave as radiation, the density of ewkons is around 300 to 30 times smaller than the density of radiation for $g$ between 1 and 10. We also found that the evolution of ewkons is consistent with dark energy throughout the universe's history: adjusting the value of $\epsilon_\infty$, ewkons currently have a barotropic parameter close to $-1$ that corresponds to the observed accelerated expansion. As we move backwards in time $w$ becomes larger and, in the large temperature regime, it takes the value 1/3 and the density behaves as $a^{-4}$. Although the density of ewkons increases in the past, it never dominates again. This last result is consistent with the hot big bang theory, which states that in the past the dynamics of the universe were dominated first by radiation and afterwards by matter.

\section{The effective scalar field potential}
\label{s.potential}

\begin{figure}
    \includegraphics[width=\linewidth]{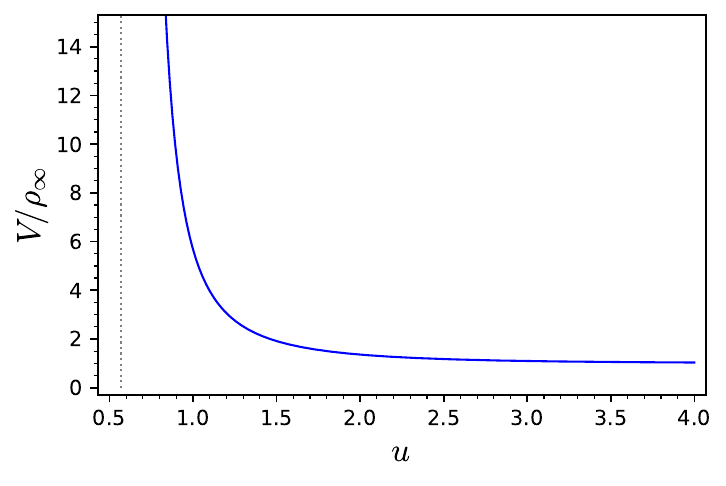}
    \caption{The potential $V$ as a function of $u=\epsilon_{\rm m}/T$; it diverges at $u\simeq 0.567$ (dotted vertical line).}
    \label{f.Vvsu}
\end{figure}

In this section we search for a scalar field effective description of the ewkons fluid. The idea of representing statistical effects with an effective potential has been applied to fermions and bosons as can be seen, for example, in Ref.\ \cite[p.\ 138]{pathria}; nevertheless it is important to keep in mind that particles are non-interacting and that, in the present cosmological context, the effective potential is the one that a scalar field should have in order to reproduce the statistical effects of ewkons. The equation of motion for the scalar field $\phi$ is
\begin{equation}\label{e.KG}
-V' \equiv -\frac{\partial V}{\partial\phi} = \ddot\phi+3H\dot\phi, 
\end{equation}
where $V(\phi)$ is the potential. The energy density and pressure of the scalar field are given as:
\begin{equation}\label{e.rhophi}
\rho_\phi = \frac{1}{2} \dot\phi^2 + V,
\end{equation}
\begin{equation}
p_\phi = \frac{1}{2} \dot\phi^2 - V.
\end{equation}
Using Eqs.\ \eqref{e.rho} and \eqref{e.p}, an expression for the potential in terms of $u$ is obtained:
\begin{align}
    V&=\frac{\rho_\phi-p_\phi}{2} \nonumber\\
    &= \rho_\infty\frac{u^{4} + 8 - 2 \, {\left(u^{3} + 2 \, u^{2} + 4 \, u + 4\right)} e^{-u}}{(u-e^{-u})^4} \xrightarrow[u \to \infty]{} \rho_\infty, \label{e.Vu}
\end{align}
where $\rho_\infty$ is given by \eqref{e.rho-inf} and from Eq.\ \eqref{e.em} it was used that the temperature is also a function of $u$:
\begin{equation}\label{e.Tu}
    T = \frac{\epsilon_\infty}{u - e^{-u}}.
\end{equation}
The shape of the potential as a function of $u$ is shown in Fig.\ \ref{f.Vvsu}.

In order to obtain $V'$ it is convenient to use the following chain rule:
\begin{equation}\label{e.Vp}
    V' = \frac{dV}{du} \frac{du}{da} \frac{da}{d\phi}.
\end{equation}
The relationship between $u$ and $a$ is given by \eqref{e.Tu} and $T=T_0/a$:
\begin{equation}\label{e.a}
    a = \alpha (u - e^{-u}),
\end{equation}
where $\alpha\equiv T_0/\epsilon_\infty$ is a constant. Consequently we have:
\begin{equation}\label{e.duda}
    \frac{du}{da} = \frac{1}{\alpha (1 + e^{-u}).}
\end{equation}
Given that $\dot{\phi}^2=\rho_\phi+p_\phi$, the derivative $da/d\phi$ can be written as
\begin{equation}\label{e.dadphi0}
    \frac{da}{d\phi} = \frac{\dot{a}}{\dot{\phi}} = \frac{a H}{\sqrt{\rho_\phi+p_\phi}},
\end{equation}
where $H=\dot{a}/a$ is the Hubble parameter. Thus we have $a$, $\rho_\phi$ and $p_\phi$ in terms of $u$. In the next subsection we show how to write also $H$ in terms of $u$ for the era dominated by dark energy. 

We have expressions \eqref{e.duda} and \eqref{e.dadphi0} for the derivatives $du/da$ and $da/d\phi$ in terms of $u$, and using them in \eqref{e.Vp} we obtain $V'$ as a function of $u$. This process can be repeated to obtain
\begin{equation}
    V'' = \frac{dV'}{du} \frac{du}{da} \frac{da}{d\phi}, 
\end{equation}
etc. The expressions obtained in this way become too large to reproduce here, but can be calculated with a computational algebra software. In particular, the limit of the derivatives of the potential when $u\rightarrow \infty$ can be computed. 
The dark energy dominated era is analyzed below.

\subsection{Dark energy dominated era}

The Friedmann equation is $H^2 = \rho_{\rm tot}/(3 m_P^2) \simeq \rho_{\rm de}/(3 m_P^2)$, where $m_P=1/\sqrt{8\pi G}$ is the reduced Planck mass and the total density $\rho_{\rm tot}$ is similar to $\rho_{\rm de}$ in the dark energy dominate era. From \eqref{e.dadphi0} we have
\begin{equation}\label{e.dadphi}
    \frac{da}{d\phi} = \frac{\alpha (u-e^{-u})}{\sqrt{3}\, m_P \sqrt{1+w}},
\end{equation}
where Eq.\ \eqref{e.a} was used for $a$, and $w$ is given by \eqref{e.wu}.

The following sequence is obtained for the derivatives of the potential:
\begin{equation}\label{e.seq}
    \begin{array}{ll}
       V^{(n)}_\infty = 0  & \text{for $n$ odd,} \\
       V^{(n)}_\infty = \frac{\rho_\infty}{6}\left(\frac{2}{m_P} \right)^n  & \text{for $n$ even,}
    \end{array}
\end{equation}
where sub-index $\infty$ indicates that the derivatives are evaluated for $u\rightarrow \infty$. The sequence was numerically checked up to $n=20$. For simplicity, we take the value of the field $\phi_\infty$, when $u\rightarrow\infty$, equal to 0. Using the Taylor expansion
\begin{equation}
    V = V_\infty + V'_\infty \phi + V''_\infty \phi^2/2! + \cdots,
\end{equation}
we obtain
\begin{equation}\label{e.Vphide}
    V(\phi) = \frac{\rho_\infty}{6} \left[5 + \cosh \left(\frac{2\phi}{m_P}\right)\right].
\end{equation}

The tracker parameter, $\Gamma = V V''/(V')^2$, is defined in order to determine whether the solution of the evolution equation of the scalar field determined by this potential is an attractor \cite[p. 211]{dimopoulos}, a condition given by $\Gamma \ge 1$. In our case, we have
\begin{equation}
    \Gamma = \frac{{\left[\cosh\left(2 \, \phi/m_P\right) + 5\right]} \cosh\left(2 \, \phi/m_P\right)}{\sinh^2(2 \, \phi/m_P)}.
\end{equation}

Since $|\cosh x| > |\sinh x|$ for any $x$, then $\Gamma>1$ and this strict inequality makes the attractor a tracker. Overall we thus expect that ours is a freezing-like scenario, where the ewkons approach a cosmological constant like behaviour in the long run \cite{Scherrer2008,Chiba2009,Gupta2015}.

\begin{figure}
    \includegraphics[width=\linewidth]{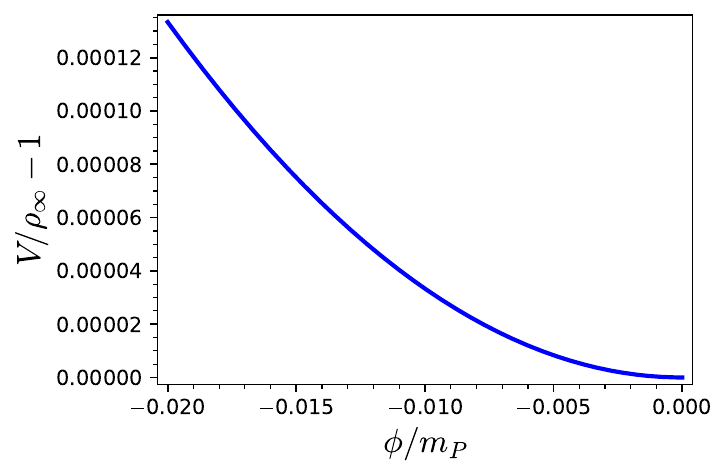}
    \includegraphics[width=\linewidth]{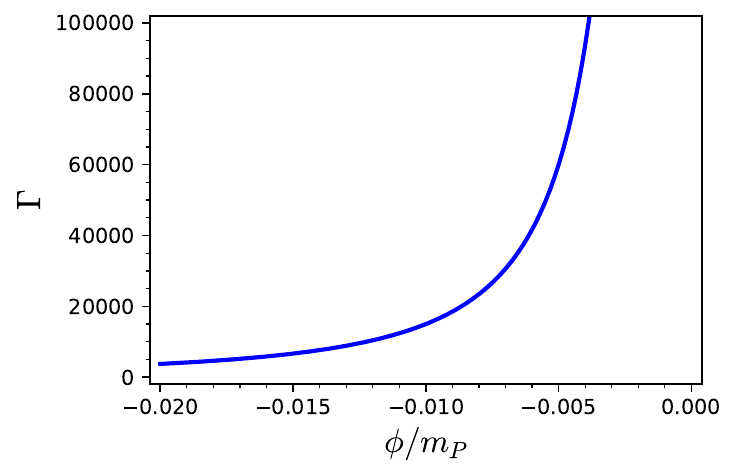}
    \caption{The potential $V$ (above), and the tracker parameter $\Gamma$ (below) as functions of the scalar field for the dark energy dominated era. $\Gamma$ diverges when $\phi$ approaches the asymptotic value $\phi_\infty=0$ at the bottom of the potential. The present value of the dimensionless scalar field $\phi_0/m_P$ is $-0.006$ for $g=1$ or $-0.02$ for $g=10$ (see Eq.\ \eqref{e.phi0de} below).}
    \label{f.potG}
\end{figure}

The shapes of the potential and the tracker parameter as functions of the scalar field are shown in Fig.\ \ref{f.potG}. The results are the same for both signs of the square root that appears in Eq.\ \eqref{e.dadphi0}, that is, for $\dot\phi$ positive or negative. Whether $\phi$ is greater than or less than $\phi_\infty=0$ as an initial condition is undetermined, as both situations are equivalent. Negative values of $\phi$ are shown in Fig.\ \ref{f.potG}.

\section{Dynamics of the scalar field $\phi$}
\label{s.scalar}

Now we calculate the time evolution of the scalar field $\phi$ using the Klein Gordon equation \eqref{e.KG} and the potential $V(\phi)$. The purpose is to verify that the potential correctly reproduces the behavior obtained in Sec.\ \ref{s.small} for small  $T$, condition that approximately overlap with the regime of dominant ewkons.

\subsection{Dark energy dominated scalar field (small temperature).}

In this subsection we consider a universe where dark energy with ewkon statistics is dominant ($\Omega_{\rm de}=1$), and analyze its dynamics with the scalar field effective description. If ewkons already account for the bulk of dark energy, the cosmology studied in this subsection would be valid from the present time throughout the far future of the Universe.
Ewkons are described by a gas of ultra relativistic particles at temperature $T$. We focus our attention at the small temperature regime: $\epsilon_{\rm m} \simeq \epsilon_\infty$, $u\simeq\epsilon_\infty/T \gg 1$, $a \gg 1$ and $|\phi|/m_P \ll 1$.

Eqs.\ \eqref{e.Vu} and \eqref{e.Vphide} give the potential $V$ as a function of $u$ or $\phi$ respectively. Approximating both equations in the present regime we have
\begin{align}
    V &= \rho_\infty (1+8/u^4) \\
    V &= \rho_\infty \left[1 + \phi^2/(3 m_P^2)\right]
\end{align}
and, combining them,
\begin{equation}\label{e.phia2}
    \frac{\phi}{m_P} = \frac{\sqrt{24}}{u^2} =\frac{\sqrt{24}\, T^2}{\epsilon_m^2}\simeq
    \frac{\sqrt{24}\, T_0^2}{\epsilon_\infty^2}\frac{1}{a^2},
\end{equation}
where both signs of the square root are possible.
From this equation, the time derivatives of the scalar field are:
\begin{align}
    \dot\phi &= -2\phi H \\
    \ddot\phi &= -2\dot\phi H
\end{align}
where in the last equation the term $-2\phi\dot{H}$ was neglected since it goes as $1/a^6$. This can be seen using the Friedmann and energy conservation equations: $\dot{H} = \dot\rho_{\rm de}/(6H m_P^2) = -(\rho_{\rm de} + p_{\rm de})/(2m_P^2) = -\dot{\phi}^2/(2m_P^2)=2\phi^2 H^2/m_P^2\sim 1/a^4$ (remember that $H\sim H_0$ up to a term proportional to $a^{-4}$). Consequently $\phi\dot{H}\sim 1/a^6$, and we conclude that $\phi$, $\dot\phi$ and $\ddot\phi$ behave as $1/a^2$. Therefore $\ddot\phi$ cannot be neglected in \eqref{e.KG} so, although the kinetic energy of the scalar field may be much less than $V$, the slow roll approximation does not fully hold.

The evolution equation for the scalar field, Eq.\ \eqref{e.KG} now in terms of $\phi$, becomes
\begin{equation}
    H\dot\phi + V'=0.
\end{equation}
Keeping terms up to order $1/a^2$, $H\simeq \sqrt{\rho_\infty}/(\sqrt{3}m_P)$, where, since $\Omega_{\rm de}=1$, we have taken $\rho_{\rm tot}=\rho_{\rm de}\simeq\rho_\infty$. Then,
\begin{equation}
    \dot\phi + \lambda \phi = 0,
\end{equation}
with
\begin{equation}\label{e.lambdadom}
    \lambda \equiv \frac{2}{m_P}\sqrt{\frac{\rho_\infty}{3}}=\sqrt{\frac{32}{3}\pi G\rho_\infty}.
\end{equation}
Then, the scalar field decays exponentially to 0:
\begin{equation}\label{e.phiscal}
\phi=\phi_0\exp{\left[-\lambda(t-t_0)\right]}
\end{equation}
with $\phi_0$ its present value. This result is a further consistent confirmation that the asymptotic value of the scalar field coincides with the minimum of the potential $\phi_\infty=0$. Replacing $a=1$ in \eqref{e.phia2}, the present value of the scalar field in $m_P$ units is given by
\begin{equation}\label{e.phi0de}
    \frac{\phi_0}{m_P} = \frac{\sqrt{24}\, T_0^2}{\epsilon_\infty^2}.
\end{equation}
Using the previously computed values of $\epsilon_\infty$, the absolute value $|\phi_0|/m_P$ is approximately equal to 0.006, for $g=1$, and 0.02, for $g=10$.

Also, from Eq.\ \eqref{e.phia2} we see that the scale factor increases exponentially:
\begin{equation} \label{e.adom}
    a = e^{\lambda(t-t_0)/2}.
\end{equation}

The density time evolution is obtained from \eqref{e.rhophi}, and the result is:
\begin{equation}
    \rho_{\rm de} = \rho_\infty\left( 1 + \left(\frac{\phi_0}{m_P}\right)^2 e^{-2\lambda(t-t_0)} \right),
\end{equation}
confirming that, at present, $\rho_0$ is approximately equal to $\rho_\infty$, as mentioned in Sec.\ \ref{s.small}.

\section{Conclusions}
\label{s.concl}

In this work we have obtained cosmological solutions for recently introduced quasi-indistinguishable particles called {\it ewkons}, which do not interact with ordinary matter (at least in the recent history of the Universe). Under the assumption of energy conservation, the cut-off energy is a time dependent quantity. These particles have a dark-energy type equation of state, which makes them a possible explanation of the accelerated expansion of the Universe. In the case of massless ewkons, the solution has the remarkable property that the presence of ewkons remains almost unnoticed until recent times, when the Universe becomes ewkon-dominated. In order to compare our proposal with current literature, we derived a scalar field effective picture of the scenario, and compared the potential obtained with other models. The potential corresponds to an effective description of these quasi-indistinguishable particles, that represents statistical effects; however we should keep in mind that the particles are non-interacting.

This is a substantially different proposal from the current literature, being based as it is on non trivial statistical assumptions. It should be explored how this might enter into the Standard Model of particles physics, and how these particles can interact with baryonic matter (even whether makes sense to assign a baryon or lepton number to them at all).


Given the great generality of this approach, early dark energy models \cite{Kojima2022} can be also included in our analysis, by choosing conveniently the ewkon parameters. However, we think desirable to better understand the theoretical basis of ewkons before looking for any scalar field effective description during the matter and radiation dominated eras. This is further required considering the lack of evidence for extensions to the $\Lambda CDM$ model \cite{Heavens2017} and the delicate observational issues involved in the current cosmological tensions \cite{Abdalla2022}. 

Finally, if we consider $\epsilon_\infty$ as the only parameter to be adjusted in our model, then we can see our approach as belonging to the {\it one-parameter dynamical dark-energy parametrizations}. However, while these models usually involve the present barotropic parameter $w_0$ as the only free parameter to be adjusted observationally with different {\it ad-hoc} functions $w(a)$ (see e.g. \cite{Yang2019} for a list of five such functions), ours predict a definite cosmological evolution for it, making it particularly interesting from a dynamical/theoretical point of view.

\section*{Acknowledgments}
This work was partially supported by Consejo Nacional de Investigaciones Científicas y Técnicas (CONICET, Argentina, PUE 22920200100016CO).

\section*{Appendix}

Here it is shown how to obtain Eq.\ \eqref{e.em} for $u=\epsilon_{\rm m}/T$ in a more general manner, without using the explicit solutions for $\rho$ and $p$ given by Eqs.\ \eqref{e.rho} and \eqref{e.p}.

It is assumed that the particles have zero (or negligible) rest energy and zero chemical potential. The pressure is
\begin{equation}\label{e.pA}
	p = \frac{T}{V} \ln \mathcal{Z} = \frac{g}{2\pi^2} \int_{0}^{\epsilon_{\rm m}} {\rm d}\epsilon \, \epsilon^2 \ln \mathcal{Z}_\mathbf{k} = \frac{g T^3}{2\pi^2} \int_{0}^{u} {\rm d}v \, v^2 \ln \mathcal{Z}_\mathbf{k}(v) = \frac{g T^3}{2\pi^2} f(u),
\end{equation}
where the variable change $v=\epsilon/T$ was used and the function $f(u)$ is defined as $f(u) = \int_{0}^{u} {\rm d}v \, v^2 \ln \mathcal{Z}_\mathbf{k}(v)$. Following a similar procedure, the density is
\begin{equation}\label{e.rhoA}
	\rho = \frac{g T^4}{2 \pi^2} \int_{0}^{u} {\rm d}v \, v^3\, \bar{n}_\mathbf{k}(v) = \frac{g T^4}{2 \pi^2} h(u),
\end{equation}
with
\begin{equation}\label{e.hA}
	h(u) = \int_{0}^{u} {\rm d}v \, v^3\, \bar{n}_\mathbf{k}(v).
\end{equation}

Replacing Eqs.\ \eqref{e.pA} and \eqref{e.rhoA} in the energy conservation equation $\dot{\rho} = 3 \frac{\dot{T}}{T} (\rho + p)$ (subscript 'de' is removed for simplicity), we obtain:
\begin{equation}\label{e.A1}
	h' \dot{u} = \frac{\dot{T}}{T} (3 f - h).
\end{equation}
In general, we have that the average number of particles in the grand canonical ensemble is $\bar{n} = T \frac{\partial\ln \mathcal{Z}}{\partial \mu}$. For the case of non-interacting particles (and $\mu=0$), we can also write the average number of particles for mode $\mathbf{k}$ as $\bar{n}_\mathbf{k} = - \frac{\partial \ln \mathcal{Z}_\mathbf{k}}{\partial (\epsilon/T)} = - \frac{\partial \ln \mathcal{Z}_\mathbf{k}}{\partial v}$; using this in \eqref{e.hA} and integrating by parts, it can be shown that
\begin{equation}\label{e.A2}
	h = -u^3 \ln \mathcal{Z}_{\rm m} + 3f,
\end{equation}
where $\mathcal{Z}_{\rm m} = \mathcal{Z}_\mathbf{k}(v=u)$. Also, using the definition of $h$, Eq.\ \eqref{e.hA}, its derivative is
\begin{equation}\label{e.A3}
	h' = u^3 \bar{n}_{\rm m} = -u^3 \frac{\partial \ln \mathcal{Z}_{\rm m}}{\partial u},
\end{equation}
where $\bar{n}_{\rm m} = \bar{n}_\mathbf{k}(v=u)$, that is, the number of particles in the highest energy level. Combining the last three equations, we have,
\begin{equation}\label{e.A4}
	- \frac{\partial \ln (\ln \mathcal{Z}_{\rm m})}{\partial u}\, {\rm d}u = \frac{{\rm  d}T}{T}.
\end{equation}
After integration, we obtain,
\begin{equation}\label{e.solA}
	\ln \mathcal{Z}_{\rm m} = \frac{c}{T},
\end{equation}
where $c$ is a constant. Using Eq.\ \eqref{e.Zewk}, we have that $\ln \mathcal{Z}_{\rm m} = -\frac{\epsilon_{\rm m}}{T} + e^{-\epsilon_{\rm m}/T}$. Considering the limit $a\rightarrow\infty$ (or $T\rightarrow 0$) we can obtain $c = -\epsilon_\infty$, and Eq.\ \eqref{e.em} is recovered.

It is interesting to notice that \eqref{e.solA} can be obtained using a different procedure. Since particles are non-interacting, each energy level can be taken as an independent system. Higher energy levels contain fewer particles, and since entropy is an extensive quantity, entropy decreases. The entropy of the highest energy level is
\begin{equation}\label{e.entropy}
	S_{\rm m} = \frac{p_{\rm m} V}{T} + \frac{U_{\rm m}}{T} = \ln \mathcal{Z}_{\rm m} + T \frac{\partial \ln \mathcal{Z}_{\rm m}}{\partial T},
\end{equation}
where $p_{\rm m}$ and $U_{\rm m}$ are the corresponding pressure and internal energy (see, for example, \cite[Ch.\ 3]{mcquarrie}). We assume that the highest energy level should be the one for which the entropy takes its minimum possible value, $S_{\rm m}=0$. With this condition, the solution of \eqref{e.entropy} is given by \eqref{e.solA}.

\bibliography{Ewkons.bib}

\end{document}